\documentclass[a4paper,12pt]{article}
\usepackage{graphicx}
\usepackage{amssymb}
\usepackage{cite}
\global\arraycolsep=2pt 
\input{epsf}
\def\Ax{\mathcal{A}}
\def\eq#1\en{\begin{equation}#1\end{equation}}
\def\s[#1,#2]{[#1\stackrel{\star}{,}#2]}
\def\pp#1{\partial_#1}
\newlength{\myVSpace}
\newcommand\xstrut{\raisebox{-.5\myVSpace}
  {\rule{0pt}{\myVSpace}}%
}
\begin{document}
\makeatletter
\def\fmslash{\@ifnextchar[{\fmsl@sh}{\fmsl@sh[0mu]}}
\def\fmsl@sh[#1]#2{%
  \mathchoice
    {\@fmsl@sh\displaystyle{#1}{#2}}%
    {\@fmsl@sh\textstyle{#1}{#2}}%
    {\@fmsl@sh\scriptstyle{#1}{#2}}%
    {\@fmsl@sh\scriptscriptstyle{#1}{#2}}}
\def\@fmsl@sh#1#2#3{\m@th\ooalign{$\hfil#1\mkern#2/\hfil$\crcr$#1#3$}}
\makeatother
\thispagestyle{empty}
\begin{titlepage}

\begin{flushright}
hep-ph/0111115v3 \\
LMU 01-12 \\
LMU-TPW 2001-10
\end{flushright}

\vspace{0.3cm}
\boldmath
\begin{center}
  {\Large {\bf The Standard Model on Non-Commutative Space-Time }}
\end{center}
\unboldmath
\vspace{0.8cm}
\begin{center}
{
{{\bf X.\ Calmet${}^{1}$, B.\ Jur\v co${}^{1}$, P.\ Schupp${}^{1}$, J.\
Wess${}^{1,2}$, M.\ Wohlgenannt${}^{1}$}}}
\end{center}
\vskip 1em
\begin{center}
${}^{1}$Sektion Physik, Universit\"at M\"unchen,\\
  Theresienstra{\ss}e 37, 80333 M\"unchen, Germany\\
${}^{2}$Max-Planck-Institut f\"ur Physik\\
        F\"ohringer Ring 6, 80805 M\"unchen, Germany\\[1em]
 \end{center}

\vspace{\fill}

\begin{abstract}
\noindent 
We consider the Standard Model on a non-commutative space and expand
the action in the non-commutativity parameter $\theta^{\mu \nu}$. No
new particles are introduced, the structure group is $SU(3)\times
SU(2)\times U(1)$.  We derive the leading order action.  At zeroth
order the action coincides with the ordinary Standard Model.  At
leading order in $\theta^{\mu\nu}$ we find new vertices which are
absent in the Standard Model on commutative space-time. The most
striking features are couplings between quarks, gluons and electroweak
bosons and many new vertices in the charged and neutral currents. We
find that parity is violated in non-commutative QCD.  The Higgs
mechanism can be applied. QED is not deformed  in the minimal
version of the NCSM to the order considered.

\end{abstract}

\end{titlepage}

\section{Introduction}
A method for implementing non-Abelian $SU(N)$ Yang-Mills theories on
non-commu\-ta\-tive space-time has recently been proposed
\cite{Madore:2000en,Jurco:2000ja,Jurco:2001my,Jurco:2001rq}. 
Previously only $U(N)$ gauge
theories were under control, and it was thus only possible to consider
extensions of the Standard Model.   Recently there has
been a lot of activity on model building. The aim of this paper is to
apply the method proposed in 
\cite{Madore:2000en,Jurco:2000ja,Jurco:2001my,Jurco:2001rq} to the
full Standard Model of particle physics \cite{Glashow:1961tr}. We
present a minimal non-commutative Standard Model with structure 
group $SU(3)_C \times SU(2)_L \times U(1)_Y$ and with the same
fields and the same number of coupling
parameters as in the Standard Model.

On a non-commutative space-time, space-time coordinates do not commute.
A particularly simple example is that of a canonical structure
\begin{equation}
  \s[x^\mu,x^\nu] \equiv x^\mu \star x^\nu- x^\nu\star x^\mu 
  =i \theta^{\mu \nu}, \label{canonical}
\end{equation}
with a constant antisymmetric matrix $\theta^{\mu \nu}$. We may think
of  $\theta^{\mu \nu}$ as a background field relative
to which directions  in space-time are distinguished.
We use the symbol~$\star$ in equation (\ref{canonical}) to denote the 
product of the non-commutative structure. We shall focus
on the case where this structure is given by 
a star product (see below), because then the discussion
of  a classical (commutative) limit is particularly transparent. 
 
Obviously, the physics on such a space-time is very different from
that on  commutative space-time. For example, Lorentz symmetry is
explicitly violated. There are several motivations to impose such a
relation, which is  reminiscent of the
non-commutative relation imposed in quantum mechanics between
coordinates and momenta. One may speculate that
space-time becomes non-commutative at very short distances when
quantum gravitation becomes relevant. We would like to point out,
however, that the non-commutativity scale could be much lower.
An example of a system  where space-time coordinates
do not commute is that of a particle in a strong
magnetic field, see e.g. \cite{Jackiw:2001dj}. 
Applying similar concepts to particle physics thus does not seem too
unnatural. Another motivation comes from string theory where non-commutative
gauge theory appears as a certain limit in the presence of a 
background field $B$ \cite{Seiberg:1999vs}.
Moreover, it is very satisfactory to understand how symmetries
can arise in a low energy theory like the Standard Model from a larger
theory which is less symmetric. Indeed the non-commutative version of
the Standard Model is Lorentz violating, but the Seiberg-Witten map
allows to understand why Lorentz symmetry is an almost exact
symmetry of Nature: the zeroth order of the theory is the Lorentz
invariant Standard Model.

If one is willing to apply the mechanism proposed in
\cite{Madore:2000en,Jurco:2000ja,Jurco:2001my,Jurco:2001rq,Bichl:2001cq} 
to the Standard Model, two problems 
have to be addressed. First, it has been claimed that in non-commutative
quantum electrodynamics charges are quantized and can only take
the values $\pm 1$ and zero \cite{Hayakawa:1999zf}.
This would indeed be a problem in view of the range of 
hypercharges in the Standard Model; see Table~\ref{tab:table1}.  
We will argue that this is really a problem  concerning the number of degrees
of freedom and will show how it can be overcome with the 
help of the Seiberg-Witten map.  
In fact the solution to the problem is closely related to the problem of 
arbitrary structure groups that we already mentioned.
Secondly, we have to deal with tensor products of gauge groups.
A no-go theorem
concerning this issue has been proposed \cite{Chaichian:2001mu}, but
we will show that this can again be dealt with 
using the methods proposed in
\cite{Madore:2000en,Jurco:2000ja,Jurco:2001my,Jurco:2001rq,Bichl:2001cq}.
We incorporate all gauge fields into one ``master field'' thereby
insuring gauge invariance of the theory.
There remains some ambiguity in the choice of kinetic terms for the
gauge potentials and for the Higgs that we shall discuss.

We expand the non-commutative action in terms of the parameter
$\theta^{\mu \nu}$. This expansion corresponds to an expansion in the
transfered momentum. It gives a low energy effective action valid
for small momentum transfer and it can be compared to the low energy
effective theory of Quantum Chromodynamics known as Chiral
Perturbation Theory (see e.g.,  \cite{deRafael:1995zv} for a review on
Chiral Perturbation Theory).  At zeroth order in $\theta^{\mu \nu}$
we recover the ordinary Standard Model.

A priori there is no reason
to  expect that $\theta^{\mu \nu}$
is constant. There is no fundamental theoretical obstacle to
formulate the theory also for non-constant
$\theta^{\mu \nu}(x)$, but we shall concentrate on the constant
case in the following for simplicity of presentation.
Up to terms involving derivatives of $\theta^{\mu \nu}(x)$, 
our leading order results are also valid for a slowly varying
space-time dependent $\theta^{\mu \nu}(x)$.
Furthermore, we should note that there are unresolved problems
with unitarity in field theories with nontrivial temporal
non-commutativity, so one has to treat that case with care. 

One of the main motivations for applying the techniques
of \cite{Madore:2000en,Jurco:2000ja,Jurco:2001my,Jurco:2001rq,Bichl:2001cq} 
to the full Standard Model is to verify that the theory is still consistent 
with the Higgs mechanism \cite{Higgs:1964pj}. 
The Higgs mechanism has previously been
discussed in the context of non-commutative Abelian gauge theory
\cite{Petriello:2001mp}.
We find that as expected the Higgs mechanism can be applied in the
non-commutative version of the Standard Model. 
The photon remains massless to all orders in the deformation parameter.
In a non-commutative setting the photon can couple to neutral particles
via a $\star$-commutator. However, in the minimal version 
of the NCSM that we present in the main part of the paper we, e.g., do not
find a vertex with two Higgs boson and one electromagnetic photon
to any order in the deformation parameter.


\section{Gauge theory on non-commutative space-time}

The subject has a long history. The idea that coordinates may not
commute can be traced back to Heisenberg. 
For an early reference on field theory on a non-commutative space,
see \cite{Snyder:1947qz}. The mathematical development of
non-commutative geometry also has a long history  \cite{connes2}.
An interpretation of the electroweak
sector in terms of non-commutative geometry has been proposed by
Connes and Lott \cite{connes1}. This is not the topic of the
present work. Neither do we consider deformations of the structure
group and corresponding quantum gauge theories, see, e.g.,
\cite{woro}. Our aim is to adapt the Standard Model 
to the situation where space-time is non-commutative.
We are in particular interested in field theory aspects of the
type of non-commutative gauge theory that has been a recent
focus of interest in string and M(atrix) theory
\cite{Seiberg:1999vs}. For a review and more references  see e.g.,\
 \cite{Douglas:2001ba}.
 
 We would like to start by briefly reviewing an intuitive approach to
 the construction of gauge theories over a given non-commutative
 structure \cite{Madore:2000en}. As such we consider a non-commutative
 associative algebra $\Ax$ whose elements we shall call
 ``functions on non-commutative space-time'' in the spirit of the
 Gel'fand-Naimark theorem. For the purposes of this
 article we shall also require that there is an invariant integral
 (trace), a well-defined classical limit and that a perturbative
 treatment of the non-commutativity is possible.  This is the case for
 the canonical structure (\ref{canonical}), which can be extended to
 the Moyal-Weyl star product defined by a formal power series
 expansion of 
 \eq (f \star g)(x)= \left.\exp\!\left(\frac{i}{2}
     \theta^{\mu \nu}\frac{\partial}{\partial x^\mu}
     \frac{\partial}{\partial y^\nu}\right) f(x) g(y)\right|_{y \to x}, 
 \en 
 together with the ordinary integral $\int d^n x\,f(x)$.  The
 latter has the property \eq \int d^n x\,(f \star g)(x) = \int d^n
 x\,(g \star f)(x) = \int d^n x\,f(x) g(x), \en as can be seen by
 partial integration.  Here $f(x)$, $g(x)$ are ordinary functions on
 $\mathbb{R}^n$ and the expansion in the star product can be seen
 intuitively as an expansion of the product in its non-commutativity.
 One should note that 
 $\mathbb{R}^n$ is only an auxiliary space needed to define the star
 product. It should not be confused with the 
 ``non-commutative space-time'' itself, which in contrast to $\mathbb{R}^n$
 does not have ``points''.  In the
 classical limit $\theta^{\mu \nu} \rightarrow 0$ we recover ordinary
 commutative space-time.

\subsection{Gauge fields on non-commutative space-time}

The construction of a gauge theory on a given 
non-commutative space can be based 
on a few basic ideas: the concept of covariant coordinates/functions, 
the requirement of locality, and gauge equivalence 
and consistency conditions. 

\subsubsection*{Non-commutative gauge transformations}

Let us consider an infinitesimal 
non-commutative local gauge 
transformation $\hat\delta$
of a fundamental matter field that carries a representation $\rho_\Psi$
\eq
\hat\delta\widehat\Psi = i \rho_\Psi(\widehat\Lambda) \star \widehat\Psi .
\label{deltapsi}
\en
In the Abelian case the representation is
fixed by the hypercharge. In the
non-Abelian case $\widehat\Psi$ is a vector, $\rho_\Psi(\widehat\Lambda)$ a
matrix whose entries are functions on non-commutative space-time and~$\star$ 
includes matrix multiplication, i.e.,
$[\rho_\Psi(\widehat\Lambda)\star \widehat\Psi]_{a}
\equiv \sum_b [\rho_\Psi(\widehat\Lambda)]_{ab} \star \widehat\Psi_b$.

The product of a field and a  coordinate, $\widehat\Psi \star x^\mu$, 
transforms
just like $\widehat\Psi$, but  the opposite product, $x^\mu\star\widehat\Psi$, 
is not a covariant object because the gauge parameter does 
not commute with $x^\mu$. In complete analogy to the covariant 
derivatives of ordinary gauge theory we thus 
need to introduce 
covariant coordinates $X^\mu  = x^\mu + \theta^{\mu\nu} \widehat A_\nu$,
where $\widehat A_\nu$ is a 
non-commutative analog of the gauge potential with the following transformation property:%
\footnote{Here and in the following we use $\theta^{\mu\nu}$ to lower 
indices, yielding expressions
that are more convenient to work with. We should note that this is 
in general only possible in
the case of constant $\theta^{\mu\nu}$.}
\eq
\hat\delta \widehat A_\mu = \pp\mu\widehat\Lambda 
+ i\s[\widehat\Lambda,\widehat A_\mu].
\label{deltaA}
\en
{}From the covariant coordinates one can construct 
further covariant objects including the
non-commutative field strength
\eq
\widehat F_{\mu\nu}  = \pp\mu\widehat A_\nu 
- \pp\nu\widehat A_\mu  -i\s[\widehat A_\mu,\widehat A_\nu],
\qquad \hat\delta\widehat F_{\mu\nu} 
= i\s[\widehat\Lambda,\widehat F_{\mu\nu}],
\en
related to the commutator of covariant coordinates, and the covariant
derivative
\eq
\widehat D_\mu \widehat\Psi = \pp\mu \widehat\Psi 
- i \rho_\Psi(\widehat A_\mu)\star\widehat\Psi,
\label{covder}
\en
related to the covariant expression 
$\rho_\Psi(X^\mu) \star \widehat\Psi - \widehat\Psi\star x^\mu$.

In the following we shall often omit the symbol~$\rho_\Psi$,
when its presence is obvious.

\subsubsection*{Locality, classical limit and Seiberg-Witten maps} 

A star product of ordinary functions $f$, $g$ can be seen as a tower
built upon its classical limit, which is determined by a Poisson
tensor $\theta^{\mu\nu}(x)$,
\eq
f \star g = f \cdot g + \frac{i}{2}\theta^{\mu\nu}(x) \pp\mu f \cdot \pp\nu g + \mathcal{O}(\theta^2)
\en with
higher order terms chosen in such a way as to yield an associative
product.  The star product is a local function of $f$, $g$, meaning
that it is a formal series 
that at each order in $\theta$ depends on $f$, $g$ and a \emph{finite}
number of derivatives of $f$ and $g$.

The non-commutative fields $\widehat A$, $\widehat \Psi$ and 
non-commutative gauge parameter $\widehat\Lambda$ can be expressed
in a similar fashion
as towers built upon the corresponding ordinary fields $A$, $\Psi$  
and ordinary gauge parameter $\Lambda$. There are so-called Seiberg-Witten
maps \cite{Seiberg:1999vs} that express the non-commutative fields and parameters as local 
functions of the ordinary fields and parameters,
\begin{eqnarray}
\widehat A_\xi[A] & = & A_\xi 
+ \frac{1}{4} \theta^{\mu\nu}\{A_\nu,\pp\mu A_\xi\} + 
\frac{1}{4} \theta^{\mu\nu}\{F_{\mu\xi},A_\nu\} +  \mathcal{O}(\theta^2),
\label{SWA}\\
\widehat \Psi[\Psi,A] & = & \Psi 
+ \frac{1}{2} \theta^{\mu\nu}\rho_\Psi(A_\nu)\pp\mu\Psi
+\frac{i}{8}\theta^{\mu\nu}[\rho_\Psi(A_\mu), \rho_\Psi(A_\nu)] \Psi +
\mathcal{O}(\theta^2),
 \label{SWPsi}\\
\widehat\Lambda[\Lambda, A] & = & \Lambda 
+ \frac{1}{4} \theta^{\mu\nu}\{A_\nu,\pp\mu \Lambda\}+ \mathcal{O}(\theta^2),
\label{SWLambda}
\end{eqnarray}
where $F_{\mu\nu} = \pp\mu A_\nu - \pp\nu A_\mu -i [A_\mu,A_\nu]$ is the
ordinary field strength.
We shall henceforth omit the explicit dependence
of the non-commutative fields and parameters on their ordinary 
counterparts with the understanding, that the hat $\,\widehat\;\,$ 
denotes non-commutative
quantities that can be expanded as local functions of 
their classical counterparts via Seiberg-Witten maps.

\subsubsection*{Gauge equivalence and consistency condition}

The Seiberg-Witten maps have the remarkable property that ordinary gauge
transformations $\delta A_\mu = \pp\mu \Lambda + i[\Lambda,A_\mu]$
and $\delta \Psi = i \Lambda\cdot \Psi$ 
induce non-commutative gauge transformations
(\ref{deltapsi}), (\ref{deltaA}) of the fields 
$\widehat A$, $\widehat \Psi$ with gauge parameter $\widehat \Lambda$ as
given above: 
\eq
\delta \widehat A_\mu = \hat\delta \widehat A_\mu,
\qquad \delta \widehat \Psi = \hat\delta \widehat \Psi. \label{gequiv}
\en
For consistency we have to require that any pair of non-commutative 
gauge parameters 
$\widehat\Lambda$, $\widehat\Sigma$ satisfy
\eq
\s[\widehat\Lambda,\widehat\Sigma] + i \delta_\Lambda \widehat\Sigma
- i \delta_\Sigma \widehat\Lambda = \widehat{[\Lambda,\Sigma]}.
\label{consistent}
\en
Since this consistency condition involves solely the gauge parameters
it is convenient to base the construction of the Seiberg-Witten map
(\ref{SWLambda}) on it. In a second step the remaining Seiberg-Witten
maps (\ref{SWA}) and (\ref{SWPsi}) can be computed from the gauge
equivalence condition (\ref{gequiv}). 
The gauge equivalence and consistency conditions do not uniquely determine
Seiberg-Witten maps. To the order considered here we have the
freedom of classical field redefinitions and non-commutative 
gauge transformations. We have used the latter freedom to
choose Seiberg-Witten maps with hermitian
$\widehat\Lambda$ and $\widehat A_\mu$.

The freedom in the Seiberg-Witten map
is essential for the renormalization of non-commutative gauge
theory~\cite{Bichl:2001cq}. The constants that parametrize the
freedom in the Seiberg-Witten map become
free running coupling constant which are determined
  by the unknown fundamental theory which is responsible for the
  non-commutative nature of space-time.
  The field redefinition freedom is also 
  important in the context of tensor products of gauge groups.

\subsection{Non-Abelian gauge groups}

The commutator 
\eq
\s[\widehat\Lambda,\widehat\Lambda'] 
= \frac{1}{2}\{\Lambda_a(x)\stackrel{\star}{,}
\Lambda'_b(x)\}[T^a,T^b] + \frac{1}{2}\s[\Lambda_a(x),\Lambda'_b(x)]\{T^a,T^b\}
\label{com}
\en
of two Lie algebra-valued non-commutative gauge parameters 
$\widehat\Lambda = \Lambda_a(x) T^a$ and $\widehat\Lambda' = \Lambda'_a(x) T^a$
does not close in the Lie algebra.
It is in general enveloping algebra-valued (it contains
products of generators), because the coefficient
$\s[\Lambda_a(x),\Lambda'_b(x)]$ of the anti-commutator of generators
$\{T^a,T^b\}$ is in general non-zero in the non-commutative case 
\cite{Madore:2000en,Jurco:2000ja}. An important exception is
$U(N)$ in the fundamental representation.
If we try, however, to construct non-commutative $SU(N)$
with Lie algebra-valued gauge parameters, we immediately face
the problem that a tracelessness condition is incompatible with (\ref{com}). 
We thus have to 
consider enveloping algebra-valued non-commutative gauge parameters
\eq
\widehat\Lambda = \Lambda^0_a(x) T^a +  \Lambda^1_{ab}(x) :T^a T^b:
+ \Lambda^2_{abc}(x) :T^a T^b T^c: + \ldots
\en
and fields. (The symbol $:\;:$ denotes some appropriate ordering of the
Lie algebra generators.) A priori we now face the problem
that we have an infinite number of parameters
$\Lambda^0_a(x)$, $\Lambda^1_{ab}(x)$, $\Lambda^2_{abc}(x)$, \ldots,
but these are not independent. They can in fact
all be expressed in terms of the right number of classical
parameters and fields via the Seiberg-Witten maps.
Similar observations and conclusions hold for the non-commutative
non-Abelian gauge fields.

\subsection{Charge in non-commutative QED}
\label{NCQED}

In non-commutative QED one faces the problem  
that the theory can apparently accommodate only
charges $\pm q$ or zero for one fixed $q$ \cite{Hayakawa:1999zf}.
We shall briefly review the problem below and will argue
that there is no such restriction in the $\theta$-expanded
approach based on Seiberg-Witten maps.
The problem (and its solution) 
is in fact related to the problem with arbitrary gauge groups 
that we discussed above: The commutation of Lie algebra-valued non-commutative
gauge parameters closes only in the fundamental representation of $U(1)$.

The only couplings of the non-com\-mu\-ta\-ti\-ve gauge boson $\widehat A_\mu$
to a matter field $\widehat \Psi$ compatible with the non-commutative
gauge transformation (\ref{deltaA}) in addition to (\ref{covder}) are
\eq
\widehat D^-_\mu \widehat\Psi^- 
= \pp\mu \widehat\Psi^- + i \widehat\Psi^-\star \widehat A_\mu,\quad
\widehat D^0_\mu \widehat\Psi^0 = \pp\mu \widehat\Psi^0,\quad
\widehat D^{0'}_\mu \widehat\Psi^{0'} = 
\pp\mu \widehat\Psi^{0'} - i \s[\widehat A_\mu,\widehat\Psi^{0'}], 
\en
with 
$\hat\delta\widehat\Psi^- = -i \widehat\Psi^-\star\widehat\Lambda$,
$\hat\delta\widehat\Psi^0 = 0$, and
$\hat\delta\widehat\Psi^{0'} = i \s[\widehat\Lambda,\widehat\Psi^{0'}]$,
respectively. 
(The latter possibility is interesting since it shows
how a neutral particle can couple to the (hyper) photon in 
a non-commutative setting \cite{Grosse:2001xz}.) 
At first sight, it thus appears that only $U(1)$ charges
$+1$, $-1$, $0$ are possible.

We should of course consider physical fields $\hat a^{(n)}_\mu(x)$. 
Let $Q$ be the generator of $U(1)$ (charge operator), $e$ 
a coupling constant and $\psi^{(n)}$ a field 
for a particle of charge $q^{(n)}$.
Then 
$A_\mu  = e Q a_\mu(x)$ and
$\hat A_\mu \star\hat\psi^{(n)} = e q^{(n)}\hat a^{(n)}_\mu(x)\star\hat\psi^{(n)}$,
since the Seiberg-Witten map $\hat A_\mu$ depends explicitly on $Q$.
In ordinary QED there is only one photon, i.e., there is 
no need for a label $(n)$ on $a_\mu$.
Here, however, we have a separate $\hat a^{(n)}_\mu$ for every charge
$q^{(n)}$ in the theory. The field strength
\eq
\widehat f^{(n)}_{\mu\nu} = \pp\mu \widehat a_\nu^{(n)} - 
\pp\nu \widehat a_\mu^{(n)} + ieq^{(n)}\s[\widehat a_\mu^{(n)},\widehat a_\nu^{(n)}]
\en
and covariant derivative
\eq
\widehat D_\mu \widehat\psi^{(n)} = \pp\mu\widehat\psi^{(n)} 
- ieq^{(n)}\widehat a_\mu^{(n)}\widehat\psi^{(n)}
\en
transform covariantly under
\eq
\hat\delta \widehat a_\mu^{(n)} = \pp\mu\widehat\lambda^{(n)} + i e
q^{(n)}\s[\widehat\lambda^{(n)},\widehat a_\mu^{(n)}],\qquad
\hat\delta \widehat\psi^{(n)} = i e q^{(n)} \widehat\lambda^{(n)}\star \widehat\psi^{(n)}.
\en
We see that the $\hat a_\mu^{(n)}$ cannot be equal to each other 
because of the non-zero $\star$-commutator in the transformation of 
$\hat a_\mu^{(n)}$. It is not 
possible to absorb $q^{(n)}$ in a redefinition of $\hat a_\mu^{(n)}$.

We can have any charge now, but
it appears that we have
too many degrees of freedom. This is not really the case, however, since all
$\hat a^{(n)}_\mu$ are local functions of the correct number of
classical gauge fields $a_\mu$ via the Seiberg-Witten map (\ref{SWA})
that, when written in terms of the physical fields, depends on $q^{(n)}$: 
\eq
\hat a^{(n)}_\xi = a_\xi 
+ \frac{e q^{(n)}}{4} \theta^{\mu\nu}\{a_\nu,\pp\mu a_\xi\} + 
\frac{e q^{(n)}}{4} \theta^{\mu\nu}\{f_{\mu\xi},a_\nu\} + \mathcal{O}(\theta^2).
\en 
In the action for the non-commutative gauge fields we now face a choice:
{}From the non-commutative point of view it appears to be natural to 
provide kinetic terms for all $\hat a^{(n)}_\mu$, even
though these fields are not really independent. This 
leads to a trace over the particles in the model and will
be discussed in Appendix~C. In the main part of the paper
we will instead make a simpler choice
for the trace that leads to minimal deviations from the
ordinary Standard Model. That choice is more
natural from the point of view that the independent degrees of freedom
are given by the $a_\mu$.  
Gauge invariance alone is not enough to favor one of the possible
choices.

\subsection{Non-commutative Yukawa couplings and Higgs}

We can generalize (\ref{SWPsi}) to the case of a field 
$\Phi$ that transforms on the left and on
the right under two arbitrary gauge groups with
corresponding gauge potentials $A_\mu$, $A'_\mu$. 
$\widehat \Phi \equiv \widehat \Phi[\Phi,A,-A']$, given by the following hybrid
Seiberg-Witten map:
\begin{eqnarray}
\widehat\Phi[\Phi,A,-A'] & = & \Phi + \frac{1}{2}\theta^{\mu\nu} A_\nu
  \Big(\pp\mu\Phi -\frac{i}{2} (A_\mu \Phi - \Phi A'_\mu)\Big)\nonumber\\
&& +\frac{1}{2}\theta^{\mu\nu} 
  \Big(\pp\mu\Phi -\frac{i}{2} (A_\mu \Phi - \Phi A'_\mu)\Big)A'_\nu
  + \mathcal{O}(\theta^2).\label{SWPhi}
\end{eqnarray}
It transforms covariantly,
\eq
\delta\widehat\Phi = i \widehat\Lambda \star\widehat\Phi - i
\widehat\Phi\star \widehat{\Lambda'},
\en
under
$\delta\Phi = i\Lambda\Phi - i\Phi\Lambda'$,
$\delta A_\nu = \pp\nu\Lambda + i[\Lambda,A_\nu]$,
$\delta A'_\nu = \pp\nu \Lambda' + i[\Lambda',A'_\nu]$.
The covariant derivative for $\widehat \Phi$ is
\eq
\widehat D_\mu \widehat\Phi = \pp\mu \widehat\Phi 
- i \widehat A_\mu\star\widehat\Phi
+ i \widehat\Phi\star\widehat{A'}_\mu.
\label{covderphi}
\en
We need the hybrid Seiberg-Witten map to construct 
gauge covariant Yukawa couplings. The  classical (``commutative'') 
Higgs $\Phi$ has $U(1)$ charge $Y=1/2$ and transforms
under $SU(2)$ in the fundamental representation. It
has no color charge. $\Phi$ obviously commutes with
the classical $U(1)$ and $SU(3)$ gauge parameters.
In the non-commutative case this is not the case
because both $\Phi$ and the parameters are functions on 
space-time and thus do not commute.
It is still true that the non-commutative $\widehat\Phi$
has overall $U(1)$ charge $Y=1/2$ and no overall color
charge, but the precise representations on the left 
(affects $A_\mu$) and on the right (affects $A'_\mu$)
are inherited from the fermions on the left and the right
of the Higgs in the Yukawa couplings.


\section{The Non-Commutative Standard Model}

The structure group of the Standard Model is 
$G_{SM}=SU(3)_C \times SU(2)_L \times U(1)_Y$.
There are several ways to deal with this
tensor product in the non-commutative case
that correspond to a freedom in the choice of
Seiberg-Witten map. The simplest, symmetric
and most natural approach is to take the classical 
tensor product and consider the whole
gauge potential $V_\mu$ of $G_{SM}$ as defined by
\begin{equation}
 {V_\nu}=g' {\cal A}_\nu(x)Y+g \sum_{a=1}^{3} B_{\nu a}(x) T^a_L
  +g_S \sum_{b=1}^{8} G_{\nu b}(x) T^b_S 
  \end{equation}
  and the commutative gauge parameter ${\Lambda}$ by
  \begin{equation}
  {\Lambda}=g' \alpha (x)Y+g \sum_{a=1}^{3} \alpha^L_{a}(x) T^a_L
  +g_S \sum_{b=1}^{8} \alpha^S_{b}(x) T^b_S,
\end{equation}
where $Y$, $T^a_L$, $T^b_S$ are the generators of
$u(1)_Y$, $su(2)_L$ and $su(3)_C$ respectively.
The non-commutative gauge parameter $\widehat \Lambda$ is
then  given via the Seiberg-Witten map by
\eq
\widehat\Lambda  =  \Lambda 
+ \frac{1}{4} \theta^{\mu\nu}\{V_\nu,\pp\mu \Lambda\}+ \mathcal{O}(\theta^2).
\en
Note that this is not equal to a naive sum of non-commutative
gauge parameters corresponding to the three factors in $G_{SM}$.
This is due to the nonlinearity of the Seiberg-Witten maps
and ultimately is a consequence of the nonlinear consistency
condition (\ref{consistent}).
The non-commutative fermion fields $\widehat \Psi^{(n)}$ corresponding to
particles labelled by $(n)$ is
\eq
\widehat \Psi^{(n)}  =  \Psi^{(n)} 
+ \frac{1}{2} \theta^{\mu\nu}\rho_{{(n)}}(V_\nu)\pp\mu\Psi^{(n)}
+\frac{i}{8}\theta^{\mu\nu}[\rho_{{(n)}}(V_\mu), \rho_{{(n)}}(V_\nu)] \Psi^{(n)} + \mathcal{O}(\theta^2).
\en
The Seiberg-Witten map for the non-commutative vector potential
$\widehat V_\mu$ yields
\begin{equation}
\widehat V_\xi  =  V_\xi 
+ \frac{1}{4} \theta^{\mu\nu}\{V_\nu,\pp\mu V_\xi\} + 
\frac{1}{4} \theta^{\mu\nu}\{F_{\mu\xi},V_\nu\} +  \mathcal{O}(\theta^2),
\end{equation}
with the ordinary field
strength $F^{\mu \nu} \equiv \partial^\mu V^\nu -\partial^\nu V^\mu
-i [ V^\mu , V^\nu ]$.  The
non-commutative field strength is
\begin{eqnarray}
\widehat F_{\mu \nu}=\partial_\mu \widehat V_\nu -
\partial_\nu \widehat V_\mu-i [\widehat V_\mu \stackrel{*}{,} \widehat
V_\nu ].
\end{eqnarray}
We have the following particle spectrum, see Table~\ref{tab:table1}:
\begin{equation}
\Psi^{(i)}_L= \left ( \matrix{ L^{(i)}_L \cr
 Q^{(i)}_L
    } \right),\qquad
 \Psi^{(i)}_R = \left ( \matrix{ e^{(i)}_R \cr
 u^{(i)}_R \cr   d^{(i)}_R
    } \right), \qquad 
    {\Phi} =
 \left(\begin{array}{c}  \phi^+ \\  \phi^0
   \end{array} \right ),
\end{equation}
where $(i)\in\{1,2,3\}$ is the generation index and 
$\phi^+$ and
$\phi^0$ are the complex scalar fields
of the scalar Higgs doublet. The non-commutative Higgs field $\widehat \Phi$
is given by the hybrid Seiberg-Witten map (\ref{SWPhi}),
\eq
\widehat\Phi = \Phi + \frac{1}{2}\theta^{\mu\nu} V_\nu
\Big(\pp\mu\Phi -\frac{i}{2} (V_\mu \Phi - \Phi V'_\mu)\Big)
+\frac{1}{2}\theta^{\mu\nu} 
\Big(\pp\mu\Phi -\frac{i}{2} (V_\mu \Phi - \Phi V'_\mu)\Big)V'_\nu
+ \mathcal{O}(\theta^2) \label{SWPhib}
\en

The non-commutative
Standard Model can now be written in a very compact way:
\begin{eqnarray}
S_{NCSM}&=&\int d^4x \sum_{i=1}^3 \overline{\widehat \Psi}^{(i)}_L \star i
\widehat{\fmslash D} \widehat \Psi^{(i)}_L
+\int d^4x \sum_{i=1}^3 \overline{\widehat \Psi}^{(i)}_R \star i
\widehat{\fmslash  D} \widehat \Psi^{(i)}_R \\
&& \nonumber -\int d^4x \frac{1}{2 g'} 
\mbox{{\bf tr}}_{\bf 1} \widehat
F_{\mu \nu} \star  \widehat F^{\mu \nu}
-\int d^4x \frac{1}{2 g} \mbox{{\bf tr}}_{\bf 2} \widehat
F_{\mu \nu} \star  \widehat F^{\mu \nu}\\
&&\nonumber
-\int d^4x \frac{1}{2 g_S} \mbox{{\bf tr}}_{\bf 3} \widehat
F_{\mu \nu} \star  \widehat F^{\mu \nu}
+ \int d^4x \bigg( \rho_0(\widehat D_\mu \widehat \Phi)^\dagger
\star \rho_0(\widehat D^\mu \widehat \Phi)            
\\ && \nonumber
- \mu^2 \rho_0(\widehat {\Phi})^\dagger \star  \rho_0(\widehat \Phi) - \lambda
\rho_0(\widehat \Phi)^\dagger \star  \rho_0(\widehat \Phi)
\star
\rho_0(\widehat \Phi)^\dagger \star  \rho_0(\widehat \Phi)   \bigg)
\\ && \nonumber
+ \int d^4x \bigg ( 
-\sum_{i,j=1}^3 W^{ij} \bigg
( ( \bar{ \widehat L}^{(i)}_L \star \rho_L(\widehat \Phi))
\star  \widehat e^{(j)}_R
+ \bar {\widehat e}^{(i)}_R \star (\rho_L(\widehat \Phi)^\dagger \star \widehat
L^{(j)}_L) \bigg )
\\ && \nonumber
-\sum_{i,j=1}^3 G_u^{ij} \bigg
( ( \bar{\widehat Q}^{(i)}_L \star \rho_{\bar Q}(\widehat{\bar\Phi}))\star  
\widehat u^{(j)}_R
+ \bar {\widehat u}^{(i)}_R \star 
(\rho_{\bar Q}(\widehat{\bar\Phi})^\dagger
\star \widehat Q^{(j)}_L) \bigg )
\\ && \nonumber
-\sum_{i,j=1}^3 G_d^{ij} \bigg
( ( \bar{ \widehat Q}^{(i)}_L \star \rho_Q(\widehat \Phi))\star  
\widehat d^{(j)}_R
+ \bar{ \widehat d}^{(i)}_R \star (\rho_Q(\widehat \Phi)^\dagger
\star \widehat Q^{(j)}_L) \bigg ) \bigg),
\end{eqnarray}
with $\bar{\Phi} = i \tau_2 \Phi^*$.
The matrices $W^{ij}$, $G^{ij}_u$ and $G^{ij}_d$ are
the Yukawa couplings. 
The gauge fields in the Seiberg-Witten maps and covariant derivatives
of the fermions terms are summarized in table~\ref{tab:table2}.
\begin{table}
\centering
  \begin{tabular}{|c|c|c|c|c|}
  \hline \addtolength{\myVSpace}{-.8cm}\xstrut
  & $SU(3)_C$ & $SU(2)_L$ & $U(1)_Y$  & $U(1)_Q$ 
   \\
   \hline
     $ e_R$
   & ${\bf 1}$
   & ${\bf 1}$ 
   & $-1$
   & $-1$  \\
  \hline \xstrut
   $ L_L=\left(\begin{array}{c} \nu_L \\ e_L \end{array} \right )$
   & ${\bf 1}$
   & ${\bf 2}$ 
   & $-1/2$
   & $\left(\begin{array}{c} 0 \\ -1 \end{array} \right )$  \\
  \hline
   $u_R$
   & ${\bf 3}$
   & ${\bf 1}$ 
   & $2/3$
   & $2/3$  \\
  \hline
   $d_R$
   & ${\bf 3}$
   & ${\bf 1}$ 
   & $-1/3$
   & $-1/3$  \\
  \hline   \xstrut
     $Q_L=\left(\begin{array}{c}  u_L \\ d_L \end{array} \right )$
   & ${\bf 3}$
   & ${\bf 2}$ 
   & $1/6$
   & $\left(\begin{array}{c} 2/3 \\ -1/3 \end{array} \right )$  \\
  \hline
  \hline\xstrut
$\Phi=\left(\begin{array}{c}  \phi^+ \\  \phi^0 \end{array} \right )$
   & ${\bf 1}$
   & ${\bf 2}$ 
   & $1/2$
   & $\left(\begin{array}{c} 1 \\ 0 \end{array} \right )$  \\
  \hline
  \hline
$B^i$
   & ${\bf 1}$
   & ${\bf 3}$ 
   & $0$
   & $(\pm 1, 0)$  \\
  \hline
$A$
   & ${\bf 1}$
   & ${\bf 1}$ 
   & $0$
   & $0$  \\
   \hline
   $G^a$
   & ${\bf 8}$
   & ${\bf 1}$ 
   & $0$
   & $0$  \\
   \hline
 \end{tabular}
 \caption{The Standard Model fields.
 The electric charge is given by the Gell-Mann-Nishijima relation
 $Q=\left(T_3+Y\right)$.
 The fields $B^i$ with $i\in \{+,-,3\}$
 denote the three electroweak gauge bosons.
 The gluons $G^i$ are in the octet representation of $SU(3)_C$.
 \label{tab:table1}}
\end{table}
The representation used in the trace of the kinetic terms for
the gauge bosons is not uniquely determined by gauge invariance
of the action. We pick the simplest choice
of a sum of traces over the $U(1)$, $SU(2)$ and $SU(3)$ 
sectors, because we are interested in a version of
the Standard Model on non-commutative space-time with
minimal modifications.\footnote{In Appendix~C we present
a different choice that is perhaps more natural from the
non-commutative point of view, with a trace over the particles in 
the Standard Model.}
In this spirit we also  take a simple choice of
representation of $Y$ of the form
 \begin{equation}
Y = \frac{1}{2} \left(\begin{array}{rr} 1 & 0 \\ 0 & -1\end{array}\right),
\end{equation}
in the definition of $\mbox{{\bf tr}}_{\bf 1}$.
The traces $\mbox{{\bf tr}}_{\bf 2}$ and trace $\mbox{{\bf tr}}_{\bf
  3}$ are the usual $SU(2)$, respectively $SU(3)$ traces. The
representations $\rho_L$, $\rho_Q$, $\rho_{\bar Q}$ of the gauge
potentials $V_\mu$, ${V'}_{\mu}$ that appear in the hybrid
Seiberg-Witten map of the Higgs are those of the fermions on the left
and right of the Higgs in the Yukawa couplings, see (\ref{SWPhi}),
\begin{eqnarray}
\rho_L(\hat\Phi[\phi,V_\mu,V'_\nu]) & = & \hat\Phi[\phi,\,-\frac{1}{2}g'{\cal A}_\mu + 
  g B^a_\mu T^a_L,\,g' {\cal A}_\nu ],\label{phiL}\\
\rho_Q(\hat\Phi[\phi,V_\mu,V'_\nu]) & = & \hat\Phi[\phi,\,\frac{1}{6} g' {\cal A}_\mu + 
  g B^a_\mu T^a_L + g_S G_\mu^a T^a_S,\,
\frac{1}{3} g' {\cal A}_\nu - 
  g_S G_\nu^a T^a_S  ],\label{phiQ}\\
\rho_{\bar Q}(\hat\Phi[\phi,V_\mu,V'_\nu]) & = & \hat\Phi[\phi,\,\frac{1}{6} g' 
  {\cal A}_\mu + g B^a_\mu T^a_L + g_S G_\mu^a T^a_S,\,-\frac{2}{3} g' {\cal A}_\nu - g_S
  G_\nu^a T^a_S].
\label{phiQbar}
\end{eqnarray}
The representation
$\rho_0$ of these gauge potentials in the kinetic term of the Higgs and in
the Higgs potential is the simplest possible one: 
\eq
\rho_0(\hat\Phi[\phi,V_\mu,V'_\nu])=\hat\Phi 
[\phi,\,\frac{1}{2} g' {\cal A}_\mu + g B^a_\mu T^a_L,\,0].
\en

There are many possibilities to choose the representations in 
the kinetic terms of the gauge bosons. 
Here, we decide to single out the choice with minimal deviations from
the Standard Model. In Appendix~C we discuss this in more
detail and present another natural
choice. Eventually physical criteria should single out the right choice.
These criteria may include, e.g., renormalization, $CPT$ invariance,
anomaly freedom, or any
kind of symmetry one might want to impose on the action. 

\begin{table}
\centering
  \begin{tabular}{|c|c|}
   \hline\addtolength{\myVSpace}{-.6cm}\xstrut
   $\Psi^{(n)}$ & $\rho_{(n)}(V_\nu)$
   \\
    \hline\hline\addtolength{\myVSpace}{-.6cm}\xstrut
   $ e_R$
   & $-g' {\cal A}_\nu(x)$   \\
  \hline\xstrut
     $ L_L=\left(\begin{array}{c} \nu_L \\ e_L \end{array} \right )$
   & $-\frac{1}{2} g' {\cal A}_\nu(x)+g B_{\nu a}(x) T^a_L$
   \\
  \hline\addtolength{\myVSpace}{-.6cm}\xstrut
   $u_R$
   & $\frac{2}{3} g' {\cal A}_\nu(x)
  +g_S  G_{\nu b}(x) T^b_S$ \\
  \hline\addtolength{\myVSpace}{-.6cm}\xstrut
   $d_R$
   & $-\frac{1}{3} g' {\cal A}_\nu(x)
  +g_S  G_{\nu b}(x) T^b_S$ \\
   \hline \xstrut
   $Q_L=\left(\begin{array}{c}  u_L \\ d_L \end{array} \right )$
   & $\frac{1}{6} g' {\cal A}_\nu(x)+g  B_{\nu a}(x) T^a_L
  +g_S G_{\nu b}(x) T^b_S$   \\
  \hline
 \end{tabular}
 \caption{The gauge fields in the Seiberg-Witten maps of the fermions and in the covariant derivatives of 
the fermions in the non-commutative Standard Model. (The symbols $T^a_L$ and
$T^b_S$ are here the Pauli and Gell-Mann matrices
respectively.) 
 \label{tab:table2}}
\end{table}
\section{The Non-Commutative Electroweak Sector} \label{ncewth}
In this section we shall apply the Seiberg-Witten map to the
electroweak non-commu\-ta\-tive Standard Model. The gauge group of the
model is $SU(3)_C \times SU(2)_L \times U(1)_Y$. The particle content
is that of the Standard Model. The matter fields and gauge fields
content is summarized in Table \ref{tab:table1}.

In the following, we shall work in the leading order of the expansion
in~$\theta$. In our convention, fields with a hat are
non-commutative whereas those without a hat are ordinary
fields.  In particular, we use the following definitions: ${\cal
  A}_\mu$ is the ordinary $U(1)_Y$ field, $B_\mu=B_\mu^i T_L^i$ are
the ordinary $SU(2)_L$  fields and $G_\mu=G_\mu^i T_S^i$ are the
ordinary $SU(3)_C$ fields. For the lepton field $L^{(i)}_L$ of the
$i$th generation which is in the fundamental representation of
$SU(2)_L$ and in the $Y$ representation of $U(1)_Y$, we have the
following expansion
\begin{eqnarray}
  \widehat L^{(i)}_L[{\cal A}, B]= L_L^{(i)} + L_L^{(i) 1}[{\cal A}, B] + {\cal
  O}(\theta^2),
\end{eqnarray}
with
\begin{eqnarray}
  L_L^{(i)1}[{\cal A}, B]
  &=&-\frac{1}{2} g'\theta^{\mu \nu} {\cal A}_\mu \partial_\nu  L_L
    -\frac{1}{2} g \theta^{\mu \nu} B_\mu \partial_\nu  L_L
   \\
  &&
  +\frac{i}{4} \theta^{\mu \nu}
  \left( g'{\cal A}_\mu +g B_\mu\right)
\left( g'{\cal A}_\nu +g B_\nu\right)
    L_L. \nonumber
\end{eqnarray}
For a right handed lepton field of the $i$th generation, one has:
\begin{eqnarray}
  \widehat e^{(i)}_R[{\cal A}]= e^{(i)}_R + e^{(i)1}_R[{\cal A}] + {\cal
  O}(\theta^2),
\end{eqnarray}
with
\begin{eqnarray}
  e^{(i)1}_R[{\cal A}]&=&-\frac{1}{2}g' \theta^{\mu \nu}
  {\cal A}_\mu \partial_\nu  e^{(i)}_R.
\end{eqnarray}
We have
\begin{eqnarray} \label{ELQ}
  \widehat Q^{(i)}_L[{\cal A}, B, G]= Q_L^{(i)} + Q_L^{(i) 1}[{\cal A}, B,G] + {\cal O}(\theta^2)
\end{eqnarray}
for a left-handed quark doublet $\widehat Q^{(i)}_L$ of the $i$th
generation, where
\begin{eqnarray} 
  Q_L^{(i)1}[{\cal A}, B,G]
  &=&-\frac{1}{2} g'\theta^{\mu \nu} {\cal A}_\mu \partial_\nu  Q_L
   -\frac{1}{2} g \theta^{\mu \nu} B_\mu \partial_\nu  Q_L
   -\frac{1}{2} g_S \theta^{\mu \nu} G_\mu \partial_\nu  Q_L \\
  &&
   +\frac{i}{4} \theta^{\mu \nu}
   \left(g' {\cal A}_\mu + g B_\mu +g_S G_\mu \right)
\left(g' {\cal A}_\nu + g B_\nu +g_S G_\nu \right)
  Q_L. \nonumber
\end{eqnarray}
For a right-handed quark e.g., $\widehat u^{(i)}_R$, we have
\begin{eqnarray} \label{ERQ}
  \widehat u^{(i)}_R[{\cal A}, G]= u_R^{(i)} + u_R^{(i) 1}[{\cal A},G]
  + {\cal O}(\theta^2),
\end{eqnarray}
\begin{eqnarray} 
  u_R^{(i)1}[{\cal A}, G]
  &=&-\frac{1}{2} g'\theta^{\mu \nu} {\cal A}_\mu \partial_\nu  u_R
  -\frac{1}{2} g_S \theta^{\mu \nu} G_\mu \partial_\nu  u_R
  \nonumber \\
  &&
   +\frac{i}{4} \theta^{\mu \nu}
   \left(g' {\cal A}_\mu +g_S G_\mu \right)
\left(g' {\cal A}_\nu +g_S G_\nu \right)
  u_R. \nonumber
\end{eqnarray}
The same expansion is obtained for a right-handed down type quark
$d_R^{(i)}$.

The field strength $\widehat F_{\mu \nu}=\partial_\mu
\widehat V_\nu - \partial_\nu \widehat V_\mu-i [\widehat V_\mu
\stackrel{*}{,} \widehat V_\nu ] $ has the following expansion:
\begin{eqnarray}
  \widehat F_{\mu \nu}&=&F_{\mu \nu}+ F^1_{\mu \nu} +{\cal O}(\theta^2),
\end{eqnarray}
with
\begin{eqnarray}
  F_{\mu \nu}&=&g'f_{\mu \nu}+g F^L_{\mu \nu}+g_S F^S_{\mu \nu},
\end{eqnarray}
where $f_{\mu \nu}$ is the field strength corresponding to the group
$U(1)_Y$, $F^L_{\mu \nu}$ that to $SU(2)_L$ and $F^S_{\mu \nu}$ that
to $SU(3)_C$. The coupling constants of the gauge groups $U(1)_Y$,
$SU(2)_L$ and $SU(3)_C$ are respectively denoted by $g'$, $g$ and $g_S$. The
leading order correction in $\theta$ is given by
\begin{eqnarray}
  F^1_{\mu \nu}&=& \frac{1}{2} \theta^{\alpha \beta} \{ F_{\mu \alpha},
  F_{\nu \beta} \} -\frac{1}{4} \theta^{\alpha \beta}
  \{ V_\alpha,(\partial_\beta+D_\beta) F_{\mu \nu} \},
\end{eqnarray}
with
\begin{eqnarray}
  D_\beta F_{\mu \nu} = \partial_\beta F_{\mu \nu} -i [V_\beta, F_{\mu \nu}].
  \end{eqnarray}

The leading order expansion for the mathematical field $V$ is given by
\begin{eqnarray}
\widehat V_\mu=V_\mu+ i\Gamma_\mu + {\cal O}(\theta^2),
\end{eqnarray}
with
\begin{eqnarray}
\Gamma_\mu & = &   i\frac{1}{4}\theta^{\alpha \beta}
     \{ g' {\cal A}_\alpha + g B_\alpha + g_S G_\alpha,
     g' \partial_\beta {\cal A}_\mu + g \partial_\beta B_\mu
+ g_S \partial_\beta G_\mu \\ && \nonumber
     + g' f_{\beta \mu} +g F^L_{\beta \mu} +g_S F^S_{\beta \mu} \}.
\end{eqnarray}

The action of the non-commutative electroweak Standard Model reads
\begin{eqnarray}
  S_{NCSM}&=&S_{Matter, leptons} + S_{Matter, quarks}
  +S_{Gauge}+S_{Higgs}+S_{Yukawa}.
\end{eqnarray}
We shall first consider the fermions (leptons and quarks). 
The fermionic matter part  is
\begin{eqnarray}
 S_{Matter, fermions}= \int d^4x \left (\sum_{f} \overline{\widehat
   \Psi}_{f L} \star
   i \fmslash D \widehat \Psi_{f L}
 + \sum_{f} \overline{\widehat \Psi}_{f R} \star i \fmslash D
 \widehat \Psi_{f R}\right), 
  \end{eqnarray}
  where $\widehat \Psi^{(f)}_{L}$ denotes the left-handed $SU(2)$ doublets
  $\widehat \Psi^{(f)}_{R}$ the right-handed $SU(2)$ singlets and the
  index $f$ runs over the three flavors. We thus have:
  \begin{equation}
    \Psi^{(1)}_{ L}=
    \left( \matrix{ {\left(\matrix{  \nu_L \cr  
    e_L}\right)}
        \cr { \matrix{\left(   \matrix{  u^r_L \cr   d^r_L}\right)
        \cr \left(   \matrix{  u^y_L \cr   d^y_L}\right)
      \cr \left(   \matrix{  u^b_L \cr   d^b_L}\right)}}}
    \right), \qquad
    \Psi^{(1)}_{R}= \left( \matrix{   e_R
        \cr   u^r_R \cr   d^r_R \cr
          u^y_R \cr   d^y_R
      \cr   u^b_R \cr   d^b_R}  \right)
    \end{equation}
    for the first generation.

We thus have
\begin{eqnarray} \label{quarkAction}
\! \! \! \! S_{Matter, fermions}&=&
 \int d^4x \bigg ( \sum_{i} \left(\bar
   L^{(i)}_{L}+  \bar L^{(i)1}_{L} \right)
\star
 i
 \left(\fmslash D^{SM} +  \fmslash \Gamma \right)
 \star
 \left(L^{(i)}_{L}+ L^{(i)1}_{L} \right)   \ \ \ \ \ \\ && + 
 \sum_{i} \left(\bar e^{(i)}_{R}+  \bar e^{(i)1}_{R} \right)
\star i
 \left(\fmslash D^{SM} +  \fmslash \Gamma \right)
 \star
 \left( e^{(i)}_{R}+ e^{(i)1}_{R} \right)
 \nonumber \bigg )  + {\cal O}(\theta^2) \\ &=& \nonumber
 \int  d^4x \sum_{i}
 \bar L^{(i)}_{L} i \fmslash{D}^{SM}  L^{(i)}_{L}\\
&&\nonumber  -\frac{1}{4} \theta^{\mu \nu}\int  d^4x \sum_{i}
 \bar L^{(i)}_{L} (g'f_{\mu \nu}+ gF^L_{\mu \nu})
 i \fmslash{D}^{SM} {L^{(i)}_{L}}
 \\ && \nonumber
 -\frac{1}{2}\theta^{\mu \nu}\int  d^4x \sum_{i}
 \bar L^{(i)}_{L} \gamma^\alpha
 (g'f_{\alpha \mu}+gF^L_{\alpha \mu}) i D^{SM}_\nu
 L^{(i)}_{L}\\
 &&\nonumber
 + \int  d^4x \sum_{i}
 \bar e^{(i)}_{R} i \fmslash{D}^{SM}  e^{(i)}_{R}
 -\frac{1}{4} \theta^{\mu \nu}\int  d^4x \sum_{i}
 \bar e^{(i)}_{R}  g'f_{\mu \nu}
 i \fmslash{D}^{SM} e^{(i)}_{R}
 \\ && \nonumber
 -\frac{1}{2}\theta^{\mu \nu}\int  d^4x \sum_{i}
\bar e^{(i)}_{R} \gamma^\alpha
 g'f_{\alpha \mu} i D^{SM}_\nu e^{(i)}_{R}  + {\cal O}(\theta^2).
\end{eqnarray}
and
\begin{eqnarray}
\! \! \! \! S_{Matter, quarks}&=&
 \int d^4x \bigg ( \sum_{i} \left(\bar
   Q^{(i)}_{L}+  \bar Q^{(i)1}_{L} \right)
\star
 i
 \left(\fmslash D^{SM} +  \fmslash \Gamma \right)
 \star
 \left(Q^{(i)}_{L}+ Q^{(i)1}_{L} \right)   \ \ \ \ \ \\ && + 
 \sum_{i} \left(\bar u^{(i)}_{R}+  \bar u^{(i)1}_{R} \right)
\star i
 \left(\fmslash D^{SM} +  \fmslash \Gamma \right)
 \star
 \left( u^{(i)}_{R}+ u^{(i)1}_{R} \right) \bigg ) \nonumber \\ \nonumber &&
+\sum_{i} \left(\bar d^{(i)}_{R}+  \bar d^{(i)1}_{R} \right)
\star i 
 \left(\fmslash D^{SM} +  \fmslash \Gamma \right)
 \star
 \left( d^{(i)}_{R}+ d^{(i)1}_{R} \right)  + {\cal O}(\theta^2)
 \\ &=& \nonumber
 \int  d^4x \sum_{i}
 \bar Q^{(i)}_{L} i \fmslash{D}^{SM}  Q^{(i)}_{L}\\
&&\nonumber  -\frac{1}{4} \theta^{\mu \nu}\int  d^4x \sum_{i}
 \bar Q^{(i)}_{L} (g'f_{\mu \nu}+ gF^L_{\mu \nu}+ g_S F^S_{\mu \nu})
 i \fmslash{D}^{SM} {Q^{(i)}_{L}}
 \\ && \nonumber
 -\frac{1}{2}\theta^{\mu \nu}\int  d^4x \sum_{i}
 \bar Q^{(i)}_{L} \gamma^\alpha
 (g'f_{\alpha \mu}+gF^L_{\alpha \mu}+g_S F^S_{\alpha \mu} ) i D^{SM}_\nu
 Q^{(i)}_{L}\\
 &&\nonumber
 + \int  d^4x \sum_{i}
 \bar u^{(i)}_{R} i \fmslash{D}^{SM}  u^{(i)}_{R}\\
&&\nonumber
 -\frac{1}{4} \theta^{\mu \nu}\int  d^4x \sum_{i}
 \bar u^{(i)}_{R}  \left ( g'f_{\mu \nu}+ g_S F^S_{\mu \nu} \right)
 i \fmslash{D}^{SM} u^{(i)}_{R}
 \\ && \nonumber
 -\frac{1}{2}\theta^{\mu \nu}\int  d^4x \sum_{i}
\bar u^{(i)}_{R} \gamma^\alpha
 \left( g'f_{\alpha \mu}+ g_S F^S_{\mu \nu} \right) i D^{SM}_\nu u^{(i)}_{R}
\\
 &&\nonumber
 + \int  d^4x \sum_{i}
 \bar d^{(i)}_{R} i \fmslash{D}^{SM}  d^{(i)}_{R}\\
&&\nonumber
 -\frac{1}{4} \theta^{\mu \nu}\int  d^4x \sum_{i}
 \bar d^{(i)}_{R}  \left ( g'f_{\mu \nu}+ g_S F^S_{\mu \nu} \right)
 i \fmslash{D}^{SM} d^{(i)}_{R}
 \\ && \nonumber
 -\frac{1}{2}\theta^{\mu \nu}\int  d^4x \sum_{i}
\bar d^{(i)}_{R} \gamma^\alpha
 \left( g'f_{\alpha \mu}+ g_S F^S_{\mu \nu} \right) i D^{SM}_\nu d^{(i)}_{R}
 + {\cal O}(\theta^2).
\end{eqnarray}
We recover the commutative Standard Model, but some new interactions
appear. The most striking feature are point-like interactions between
gluons, electroweak bosons and quarks.  For the gauge part of the
action, one finds
\begin{eqnarray}
  S_{gauge}&=&-\int d^4x \frac{1}{2 g'} 
\mbox{{\bf tr}}_{\bf 1} \widehat
F_{\mu \nu} \star  \widehat F^{\mu \nu} 
-\int d^4x \frac{1}{2 g} \mbox{{\bf tr}}_{\bf 2} \widehat
F_{\mu \nu} \star  \widehat F^{\mu \nu}
-\int d^4x \frac{1}{2 g_S} \mbox{{\bf tr}}_{\bf 3} \widehat
F_{\mu \nu} \star  \widehat F^{\mu \nu}
\nonumber\\ 
 &=&-\frac{1}{4} \, \int d^4x \, f_{\mu \nu} f^{ \mu \nu}
 \nonumber \\ \nonumber &&
-\frac{1}{2} \, {\rm Tr} \int d^4x \, F^L_{\mu \nu} F^{L \mu \nu}
 -g \, \theta^{\mu
\nu} \, {\rm Tr} \int d^4x \, F^L_{\mu \rho} F^L_{\nu \sigma} F^{L \rho
\sigma} \\ \nonumber &&
-\frac{1}{2} \, {\rm Tr}  \int d^4x \, F^S_{\mu \nu} F^{S \mu \nu}
+ \frac{1}{4}g_S \, \theta^{\mu \nu} \, {\rm Tr} \int d^4x \, F^S_{\mu \nu}
F^S_{\rho \sigma} F^{S \rho \sigma}\\
&&  - g_S \, \theta^{\mu
\nu} \, {\rm Tr} \int d^4x \, F^S_{\mu \rho} F^S_{\nu \sigma} F^{S \rho
\sigma} + {\cal O}(\theta^2).
\end{eqnarray}
The coefficients of the triple vertex in the $U(1)$ sector are
also different from plain NCQED with a single electron. These
coefficients depend on the representation we are choosing for the
$Y$ in the kinetic terms. For the simple choice that we
have taken ${\rm {\bf tr_1}} Y^3 =0$ and this coefficient is zero. 
Note that a term
\begin{eqnarray}
+ \frac{1}{4}g \, \theta^{\mu \nu} \, {\rm Tr} \int d^4x
\, F^L_{\mu \nu}
F^L_{\rho \sigma} F^{L \rho \sigma}
\end{eqnarray}
vanishes, the trace over the three Pauli matrices yields $2i
\epsilon^{abc}$ and the sum $\epsilon^{abc} F^{bL}_{\rho \sigma} F^{c
  L \rho \sigma}$ vanishes. 
Note that because the trace over $\tau^3 \tau^3 \tau^3$ vanishes,
there is also no cubic self-interaction term for the electromagnetic photon
coming from the $SU(2)$ sector. Limits on
non-commutative QED found from triple photon self-interactions 
do therefore not apply for the minimal non-commutative Standard Model.

As in the usual commutative Standard Model, the Higgs mechanism can be
applied to break the $SU(2)_L\times U(1)_Y$ gauge symmetry and thus to
generate masses for the electroweak gauge bosons.  The non-commutative
action for a scalar field $\phi$ in the fundamental representation of
$SU(2)_L$ and with the hypercharge $Y=1/2$ reads:
\begin{eqnarray}
  S_{Higgs}&=&\int d^4x \bigg (
  \rho_0\left( D_\mu \widehat \Phi \right)^\dagger \star\rho_0 \left( D^\mu  \widehat
    \Phi \right)\\
\nonumber & - & 
\mu^2  \rho_0(\widehat \Phi)^\dagger \star \rho_0( \widehat \Phi) -
\lambda ( \rho_0(\widehat \Phi)^\dagger \star \rho_0( \widehat \Phi)) \star 
(\rho_0( \widehat \Phi)^\dagger \star \rho_0(\widehat \Phi)) \bigg ). 
  \end{eqnarray}
 In the leading order of the expansion in $\theta$, we obtain:
 \begin{eqnarray}
   S_{Higgs}&=& \int d^4x\Bigg( (D^{SM}_\mu\phi)^\dagger D^{SM \mu}\phi
   -\mu^2 \phi^\dagger \phi
-\lambda (\phi^\dagger \phi) (\phi^\dagger \phi) \Bigg)
   \\ \nonumber &&
   +
   \int d^4x \Bigg ( (D^{SM}_\mu\phi)^\dagger
   \left( D^{SM \mu}\rho_0(\phi^1) + \frac{1}{2}
   \theta^{\alpha \beta} \partial_\alpha V^{\mu} \partial_\beta \phi 
 + \Gamma^\mu \phi \right)
\\ \nonumber && +
\left(D^{SM}_\mu \rho_0 (\phi^1) + \frac{1}{2}
   \theta^{\alpha \beta} \partial_\alpha V_\mu \partial_\beta \phi 
 + {\Gamma_\mu} \phi \right)^\dagger D^{SM \mu}\phi
\\ \nonumber &&
+\frac{1}{4} \mu^2
\theta^{\mu \nu} \phi^\dagger (g' f_{\mu \nu} + g F^L_{\mu \nu}) \phi
-  \lambda i \theta^{\alpha \beta}
\phi^\dagger \phi (D^{SM}_\alpha \phi)^\dagger (D^{SM}_\beta \phi)
\Bigg) + {\cal O}(\theta^2),
  \end{eqnarray}
  with
  \begin{eqnarray}
    \Gamma_\mu=-i V^1_\mu=i\frac{1}{4}
    \theta^{\alpha \beta}
    \{g'{\cal A}_\alpha+g B_\alpha,g'\partial_\beta {\cal A}_\mu
    +g \partial_\beta B_\mu + g' f_{\beta \mu} + g F^L_{\beta \mu} \}
    \end{eqnarray}
    and
    \begin{equation}
    \rho_0(\hat \Phi)=\phi+\rho_0(\phi^1)+\mathcal{O}(\theta^2),
    \end{equation}
    where
    \begin{eqnarray}
      \rho_0(\phi^1)=-\frac{1}{2} \theta^{\alpha \beta}
      (g'{\cal A}_\alpha+g B_\alpha) \partial_\beta \phi
      +i\frac{1}{4} \theta^{\alpha \beta}
      (g'{\cal A}_\alpha+g B_\alpha) (g'{\cal A}_\beta+g B_\beta) \phi.
\end{eqnarray}
For $\mu^2<0$ the $SU(2)_L\times U(1)_Y$ gauge symmetry is
spontaneously broken to $U(1)_{Q}$, which is the gauge group
describing the electromagnetic interactions. We have a gauge freedom
and take the so-called unitarity gauge
  \begin{eqnarray}
\phi &=&\left(\begin{array}{c} 0 \\
     \eta +v \end{array} \right ){1\over \sqrt{2}},
\end{eqnarray}
where $v$ is the vacuum expectation value. Since the leading order of
the expansion of the non-commutative action corresponds to the
Standard Model action, the Higgs mechanism generates masses for
electroweak gauge bosons:
 \begin{eqnarray}
 M_{W^\pm}=\frac{g v}{2} \ \ \ \mbox{and} \ \ \
 M_Z=\frac{\sqrt{g^2+g'^2}}{2} v,
 \end{eqnarray}
 where the physical mass eigenstates $W^\pm$, $Z$ and $A$ are as
 usual defined by:
\begin{equation}
   W^\pm_\mu=\frac{B^1_\mu \mp i B^2_\mu}{\sqrt{2}}, \quad
   Z_\mu=\frac{-g'{\cal A}_\mu+gB^3_\mu}{\sqrt{g^2+g'^2}}
\quad \mbox{and} \quad
   {A}_\mu=\frac{g{\cal A}_\mu+g' B^3_\mu}{\sqrt{g^2+g'^2}}.
   \end{equation}
   The Higgs mass is then given by $m^2_\eta=-2\mu^2$. Rewriting the
   term $\Gamma_\mu$ in terms of the mass eigenstates, using
\begin{eqnarray}
   B^3_\mu=\frac{g Z_\mu+g' A_\mu}{\sqrt{g^2+g'^2}} \ \ \ \mbox{and} \ 
\ \
   {\cal A}_\mu=\frac{g A_\mu-g' Z_\mu}{\sqrt{g^2+g'^2}},
\end{eqnarray}
one finds that besides the usual Standard Model couplings, numerous
new couplings between the Higgs boson and the electroweak gauge bosons
appear. 
We note that the non-commutative version
of the Standard Model is also compatible with the alternative to the
Higgs mechanism proposed in \cite{Calmet:2000th}.

The Yukawa couplings can then generate masses for the fermions, one has:
\begin{eqnarray}
S_{Yukawa}&=&\int d^4x \bigg ( 
-\sum_{i,j=1}^3 W^{ij} \bigg
( ( \bar{ \widehat L}^{(i)}_L \star \rho_L(\widehat \Phi))\star  
\widehat e^{(j)}_R
+ \bar {\widehat e}^{(i)}_R \star (\rho_L(\widehat \Phi)^\dagger \star \widehat
L^{(j)}_L) \bigg )
\\ && \nonumber
-\sum_{i,j=1}^3 G_u^{ij} \bigg
( ( \bar{\widehat Q}^{(i)}_L \star \rho_{\bar{Q}}(\widehat{\bar{\Phi}}))\star  
\widehat u^{(j)}_R
+ \bar {\widehat u}^{(i)}_R \star (\rho_{\bar{Q}}(\widehat{\bar \Phi})^\dagger
\star \widehat Q^{(j)}_L) \bigg )
\\ && \nonumber
-\sum_{i,j=1}^3 G_d^{ij} \bigg
( ( \bar{ \widehat Q}^{(i)}_L \star \rho_Q(\widehat  \Phi))\star  
\widehat d^{(j)}_R
+ \bar{ \widehat d}^{(i)}_R \star (\rho_Q(\widehat \Phi)^\dagger
\star \widehat Q^{(j)}_L) \bigg ) \bigg),
\end{eqnarray}
with $\widehat\Phi[\Phi,V,V']$ as given in (\ref{phiL})-(\ref{phiQbar}).
The sum runs over the different generations. The leading order
expansion is
\begin{eqnarray}
S_{Yukawa}&=&S^{SM}_{Yukawa}
- \int d^4x \bigg ( \sum_{i,j=1}^3 W^{ij} \bigg (
( \bar L^{i}_L \phi)   e^{1 j}_R +
( \bar L^{i}_L \rho_L(\phi^1))   e^{j}_R \\
&+& ( \bar L^{1 i}_L \phi)   e^{j}_R +
i\frac{1}{2}\theta^{\alpha \beta} \partial_\alpha L^i_L \partial_\beta 
\phi e^{j}_R
+ \bar e^{i}_R (\phi^{ \dagger} L^{1 j}_L)\nonumber\\
&+& \bar e^{i}_R (\rho_L(\phi^1)^{ \dagger} L^{j}_L)+
\bar e^{1 i}_R (\phi^{\dagger} L^{j}_L) +
i\frac{1}{2}\theta^{\alpha \beta}
\partial_\alpha e^{i}_R \partial_\beta \phi^\dagger L^j_L
 \bigg)\nonumber
 \\ - &&
\sum_{i,j=1}^3 G_u^{ij} \bigg (
( \bar Q^{i}_L \bar \phi)   u^{1 j}_R +
( \bar Q^{i}_L \rho_{\bar{Q}}(\bar \phi^1))   u^{j}_R +
( \bar Q^{1 i}_L \bar \phi)   u^{j}_R \nonumber\\
&+&
i\frac{1}{2}\theta^{\alpha \beta} \partial_\alpha Q^i_L \partial_\beta
\bar \phi u^{j}_R
+ \bar u^{i}_R (\bar \phi^{\dagger} Q^{1 j}_L)+
\bar u^{i}_R (\rho_{\bar{Q}}(\bar \phi^{1})^ \dagger Q^{j}_L)\nonumber\\
&+&
\bar u^{1 i}_R (\bar \phi^{\dagger} Q^{j}_L)
+
i\frac{1}{2}\theta^{\alpha \beta}
\partial_\alpha u^{i}_R \partial_\beta \bar \phi^\dagger Q^j_L
 \bigg)\nonumber
 \\ - &&
\sum_{i,j=1}^3 G_d^{ij} \bigg (
( \bar Q^{i}_L \phi)   d^{1 j}_R +
( \bar Q^{i}_L \rho_Q(\phi^1))   d^{j}_R +
( \bar Q^{1 i}_L \phi)   d^{j}_R\nonumber\\
&+&
i\frac{1}{2}\theta^{\alpha \beta} \partial_\alpha Q^i_L \partial_\beta 
\phi d^{j}_R
+ \bar d^{i}_R (\phi^{\dagger} Q^{1 j}_L)+
\bar d^{i}_R (\rho_Q(\phi^1)^{ \dagger} Q^{j}_L)\nonumber\\
&+&
\bar d^{1 i}_R (\phi^{\dagger} Q^{j}_L) +
i\frac{1}{2}\theta^{\alpha \beta}
\partial_\alpha \bar d^{i}_R \partial_\beta \phi^\dagger Q^j_L
 \bigg) \bigg)+ {\cal O}(\theta^2),\nonumber
\end{eqnarray}
where $L^i_L$ stands for a left-handed leptonic doublet of the $i$th
generation, $e^i_R$ for a leptonic singlet of the $i$th generation,
$Q^i_L$ for a left-handed quark doublet of the $i$th generation,
$u^i_R$ for a right-handed up-type quark singlet of the $i$th and
$d^i_R$ stands for a right-handed down-type quark singlet of the $i$th
generation. We used
\begin{equation}
\rho(\Phi)=\phi+\rho(\phi^1)+\mathcal{O}(\theta^2),
\end{equation}
where $\rho$ stands for $\rho_L$, $\rho_Q$ and $\rho_{\bar Q}$, respectively. 
$\rho(\phi^1)$ is given by (\ref{SWPhi}),
\begin{equation}
\rho(\phi^1)=\frac{1}{2} \theta^{\mu\nu}\rho(V_\nu)
\Big(\pp\mu\phi -\frac{i}{2} \rho(V_\mu) \phi + \frac{i}{2} \phi \,
\rho(V'_\mu)\Big) + \frac{1}{2} \theta^{\mu\nu}
\Big(\pp\mu\phi -\frac{i}{2} \rho(V_\mu) \phi + \frac{i}{2} \phi \,
\rho(V'_\mu)\Big)\rho(V'_\nu).
\end{equation}
Once again we recover the Standard Model, but some new
interactions arise. The Yukawa coupling matrices can be diagonalized
using biunitary transformations. We thus obtain a Cabibbo Kobayashi
Maskawa matrix in the charged currents, as in the Standard Model and
as long as right-handed neutrinos are absent, we do not predict lepton
flavor changing currents. We give the Lagrangian for the charged
currents in Appendix~A and that for the neutral currents in
Appendix~B. Clearly, flavor physics is much richer than in the
Standard Model on a commutative space. 

\section{Non-Commutative Quantum Chromodynamics}

The method developed in \cite{Jurco:2001rq} has been applied to
non-commutative Quantum Chromodynamics NCQCD already
\cite{Carlson:2001sw}. But the authors of \cite{Carlson:2001sw} have
only considered the gauge group $SU(3)_C$ instead of $SU(3)_C \times
SU(2)_L \times U(1)_Y$ which is the relevant gauge group to describe
charged quarks. Our results are thus different since the quarks are
not only in the fundamental representation of $SU(3)$ but they are
also charged under $SU(2)_L \times U(1)_Y$. This implies in particular
that parity is broken in NCQCD in the leading order of the expansion
in $\theta$ as the left-handed quarks are charged under $SU(2)_L$. One thus
have to treat the right-handed and left-handed quarks separately. The
expansion for the non-commutative quarks is thus of the form
(\ref{ELQ}) for a left-handed quark $Q_L$ and of the form (\ref{ERQ})
for a right-handed quark $Q_R$. The non-commutative action has
actually already been given previously in equation (\ref
{quarkAction}), although it appears in a hidden fashion.  

\section{Discussion of the model}

We have shown in section \ref{ncewth} that the commutative electroweak
Standard Model comes out as the zeroth order of the expansion in
$\theta^{\mu \nu}$ of the action of the non-commutative Standard
Model (NCSM).  Although we have considered a minimal non-commutative
Standard Model, there is a basic difference between the commutative
and non-commutative versions: in the non-commutative model, the
different interactions cannot be considered separately as the
master field $V_\mu$ which is a superposition of the different
gauge fields has to be introduced. In the leading order of the
expansion in $\theta$, we find that the gauge bosons of the different gauge
groups decouple. But, because the quarks are charged under $SU(3)_C$
as well as under $SU(2)_L \times U(1)_{Y}$ some new vertices appear
where the gauge bosons of different gauge groups are connected to the
quarks.  In the minimal non-commutative Standard Model, a kind of
mixing or unification between all the interactions appears as we have
vertices where e.g.,  $SU(3)_C$ gauge bosons couple to the $U(1)_{Y}$
gauge boson and to quarks. This type of unification implies
that parity is broken in NCQCD.

Up to the order considered 
we do not find couplings of neutral particles like the Higgs boson to
the electromagnetic photon in the minimal version of the NCSM.
We also find new vertices in the pure gauge sector. In contradiction
with naive expectations the $U(1)_Y$ gauge boson has not a
self-interacting vertex to the order considered, but 
one finds vertices with five and six gauge
bosons for the gauge group $SU(3)_C$ and $SU(2)_L$.

All the important features of the ordinary Standard Model can be
implemented in the model, in particular the Higgs mechanism and the
Yukawa sector. Biunitary transformations can be applied to diagonalize
the matrices of Yukawa couplings.

Recently a model based on the gauge group $U(3) \times U(2) \times
U(1)$ was proposed \cite{Chaichian:2001py}. This model involves a clever
extra Higgs mechanism to deal with the problems of charge quantization
and tensor products, but it contains two gauge bosons which are
not present in the usual Standard Model. What we are doing is
fundamentally different as we are considering the Standard Model gauge
group $SU(3) \times SU(2) \times U(1)$ directly.  We thus have proposed a
\emph{minimal} non-commutative extension of the Standard Model.

We have presented the first order expansion in $\theta^{\mu \nu}$ of
the non-commutative Standard Model, which only represents a low energy
effective theory. The limits that can be found in the literature on
the combination $\Lambda \theta$ are based on the assumption that
$\theta^{\mu \nu}$ is constant \cite{limits}, clearly the limits are
much weaker if the assumption is relaxed. As in the case of chiral
perturbation theory, the effects are expected to be small for light
particles. But, they could be sizable for heavy particles. In
particular it is conceivable that a phase transition occurs a high
energy, Nature could be non-commutative above that scale but
commutative under the scale of this phase transition.

Clearly the Standard Model on a non-commutative space-time predicts a
lot of new physics beyond the Standard Model. In particular as we have
seen, we expect the charged and neutral currents to be considerably
affected by non-commutative physics. The extraction of the CKM matrix
elements and in particular of the phase at the origin of
$CP$-violation would be strongly influenced by that type of new
physics. One expects that the effects should become larger with the
mass of the decaying particle, especially if a phase transition
exists. This might explain why the Standard Model on a commutative
space can accommodate accurately $CP$-violation in the Kaon system
although large non-commutative effects could show up in e.g. the
$B$-meson system.

High energy cosmic rays are also a place to probe non-commutative
physics, it has been proposed by S. Coleman and S. Glashow
\cite{Coleman:1999ti} that a violation of Lorentz invariance could
explain this phenomenon.

There are good reasons to think that our model is renormalizable to
all orders in the coupling constants and in $\theta$:
A study in the framework of non-commutative quantum electrodynamics
\cite{Bichl:2001cq} has shown that the photon self-energy is
renormalizable to all order. But, a proof of the renormalizability of
our model is still to be furnished.  
 The problem of
ultra-violet and infra-red mixing which plagues non-commutative
quantum field theories \cite{UVIR}, should be reconsidered in the framework of
the Seiberg-Witten expansion used in our approach \cite{Bichl:2001nf}.
 Note that the
ultra-violet and infra-red mixing is absent in the case of
$\Phi^4$-theory on a fuzzy sphere \cite{Chu:2001xi}, where the
quantization has been performed via path integrals.

\section{Conclusions}

We have considered the minimal non-commutative extension of the
Standard Model (NCSM) and have calculated the first order expansion of the
model in $\theta^{\mu \nu}$. This required to solve two problems: the
$U(1)$ charge quantization and the application of the Seiberg-Witten
method to a tensor product of groups. The trace over the 
field strength has to be defined properly. We obtain a low energy
effective theory valid for small transfered momentum, in that sense it
is the analog of Chiral Perturbation Theory for Quantum
Chromodynamics.  The zeroth order expansion is the commutative
Standard Model. The model has the same number of free coupling
constants and fields as the usual Standard Model.

We find that the most striking feature of the model is a new type of
unification as all interactions have to be considered simultaneously.
We have found that the Higgs boson does not couple to the electroweak
photon in the minimal NCSM and that new effects in the charged and neutral currents are
expected.  This will affect the extraction of the CKM matrix
parameters and in particular of the $CP$-violating phase. Neutral
decays of heavy particle, e.g., of the $b$ and $t$-quarks might also
reveal the non-commutative nature of space-time. New vertices appear
in QCD.  We find a point-like interaction between two quarks a gluon
and a photon, thus opening new decay modes for hadrons. Parity is
violated in the leading order of the expansion in the non-commutative
parameter $\theta$.


The non-commutative Standard Model represents a very natural extension
of the Standard Model, it could improve some of its problems like
naturalness and the so-called hierarchy problem and it represents a
natural attempt to include effects of quantum gravity in particle
physics.

\section*{Acknowledgements}

We would like to thank P.~Aschieri, W.~Behr and  L. M\"oller 
for many helpful discussions. We would in particular like
to thank W. Behr for pointing out subtleties concerning the gauge invariance
of the Yukawa couplings.
P.S. gratefully acknowledges the hospitality of Lawrence Berkeley National Laboratory
and helpful discussions with B. Zumino and his group.
X.C. acknowledges partial support by the Deutsche
Forschungsgemeinschaft, DFG-No. FR 412/27-2.

\section*{Appendix A: Charged currents} \label{appendixB}

In this section we give the explicit formulas for the electroweak
charged currents in the leading order of the expansion in $\theta$.
\begin{eqnarray}
  {\cal L}&=&\left (\matrix{ \bar u & \bar c  & \bar t} \right)_L V_{CKM} J_1
  \left(\matrix{ d \cr s  \cr b } \right)_L +
  \left(\matrix{ \bar d & \bar s  & \bar b } \right)_LV^\dagger_{CKM} J_2
  \left (\matrix{ u \cr c  \cr t} \right)_L,
\end{eqnarray}
with
\begin{eqnarray}
  J_1&=& \frac{1}{\sqrt{2}} g \fmslash W^+ +  (\frac{1}{2} \theta^{\mu \nu}
  \gamma^\alpha + \theta^{\nu \alpha}
  \gamma^\mu ) 
   \\ && \nonumber
   \Bigg( \Big( -\frac{\sqrt{2}}{4} Y g' g (\cos{\theta_W} \partial_\mu A_\nu
-\cos{\theta_W} \partial_\nu A_\mu
-\sin{\theta_W} \partial_\mu Z_\nu
+\sin{\theta_W} \partial_\nu Z_\mu) W^+_\alpha \Big)
\\ && \nonumber
+g \frac{\sqrt{2}}{8}  \bigg( \partial_\mu W^+_\nu -  \partial_\nu W^+_\mu
    \\ && \nonumber  
  -2 i g \left ( \cos{\theta_W} Z_\mu W^+_\nu +
  \sin{\theta_W} A_\mu W^+_\nu
  - \cos{\theta_W} W^+_\mu Z_\nu
  - \sin{\theta_W} W^+_\mu A_\nu \right) \bigg ) \cdot
 \\ && \nonumber
 (-2 i \partial_\alpha + 2 Y g' \sin{\theta_W} Z_\alpha
 - 2 Y g' \cos{\theta_W} A_\alpha  + g  \cos{\theta_W} Z_\alpha +
 g \sin{\theta_W} A_\alpha )
 \\ && \nonumber
 - \frac{\sqrt{2}}{8} g^2 \Big( \cos{\theta_W} \partial_\mu Z_\nu
 - \cos{\theta_W} \partial_\nu Z_\mu
 + \sin{\theta_W} \partial_\mu A_\nu
- \sin{\theta_W} \partial_\nu A_\mu
\\ && \nonumber
-2ig (W^+_\mu W^-_\nu - W^+_\nu W^-_\mu)\Big) W^+_\alpha
 \Bigg)
\end{eqnarray}
and
\begin{eqnarray}
  J_2&=& \frac{1}{\sqrt{2}} g \fmslash W^- +  (\frac{1}{2} \theta^{\mu \nu}
  \gamma^\alpha + \theta^{\nu \alpha}
  \gamma^\mu ) 
   \\ && \nonumber
   \Bigg( \Big( -\frac{\sqrt{2}}{4} Y g' g (\cos{\theta_W} \partial_\mu A_\nu
-\cos{\theta_W} \partial_\nu A_\mu
-\sin{\theta_W} \partial_\mu Z_\nu
+\sin{\theta_W} \partial_\nu Z_\mu) W^-_\alpha \Big)
\\ && \nonumber
+g \frac{\sqrt{2}}{8}  \bigg( \partial_\mu W^-_\nu -  \partial_\nu W^-_\mu
    \\ && \nonumber  
  -2 i g \left ( \cos{\theta_W}   W^-_\mu Z_\nu +
  \sin{\theta_W} W^-_\mu A_\nu
  - \cos{\theta_W}  Z_\mu W^-_\nu
  - \sin{\theta_W} A_\mu W^-_\nu \right) \bigg ) \cdot
 \\ && \nonumber
 (-2 i \partial_\alpha + 2 Y  g' \sin{\theta_W} Z_\alpha
 - 2 Y g' \cos{\theta_W} A_\alpha  - g  \cos{\theta_W} Z_\alpha -
 g \sin{\theta_W} A_\alpha )
 \\ && \nonumber
 - \frac{\sqrt{2}}{8} g^2 \Big( \cos{\theta_W} \partial_\mu Z_\nu
 - \cos{\theta_W} \partial_\nu Z_\mu
 + \sin{\theta_W} \partial_\mu A_\nu
- \sin{\theta_W} \partial_\nu A_\mu
\\ && \nonumber
-2ig (W^+_\mu W^-_\nu - W^+_\nu W^-_\mu)\Big) W^-_\alpha
 \Bigg).
 \end{eqnarray}
 Note that we have not included the interactions with the gluons in
 the ``electroweak'' charged currents. These formulas can be further simplified
 using the identity $g' \cos\theta_W = g \sin \theta_W$.

 \section*{Appendix B: Neutral currents}

In this appendix we give the explicit formula for the neutral current
in the leading order of the expansion in $\theta$.
\begin{eqnarray}
{\cal L}_{nc}&=&{\cal L}^{SM}_{nc} - i \frac{1}{2}
\sum_i\bar u^{(i)}_L \left(\frac{1}{2} \theta^{\mu \nu}
  \gamma^\alpha + \theta^{\nu \alpha}
  \gamma^\mu \right) \\ && \nonumber
\Bigg( \bigg(\cos{\theta_W} \partial_\mu A_\nu
-\cos{\theta_W} \partial_\nu A_\mu
-\sin{\theta_W} \partial_\mu Z_\nu
+\sin{\theta_W} \partial_\nu Z_\mu \bigg)
\\ && \nonumber
\bigg( g' Y \partial_\alpha - i Y^2 g'^2 \cos{\theta_W} A_\alpha
+ i Y^2 g'^2 \sin{\theta_W} Z_\alpha - i \frac{1}{2} Y g' g  \cos{\theta_W} Z_\alpha
\\ && \nonumber
- i \frac{1}{2} Y g' g  \sin{\theta_W} A_\alpha 
\bigg)
+ \frac{1}{2}\bigg( \cos{\theta_W} \partial_\mu Z_\nu
 - \cos{\theta_W} \partial_\nu Z_\mu
 + \sin{\theta_W} \partial_\mu A_\nu
- \sin{\theta_W} \partial_\nu A_\mu
\\ && \nonumber
-2ig (W^+_\mu W^-_\nu - W^+_\nu W^-_\mu) \bigg) \bigg (
g \partial_\alpha -i Y g' g \cos{\theta_W}A_\alpha
\\ && \nonumber
 + i Y g' g  \cos{\theta_W} Z_\alpha
 - \frac{1}{2} i g^2 \cos{\theta_W}Z_\alpha
 - \frac{1}{2} i g^2 \sin{\theta_W}A_\alpha
\bigg)
\\ \nonumber &&
-   \frac{i}{2} g^2\bigg(\partial_\mu W^+_\nu -\partial_\nu W^+_\mu
\\ \nonumber &&
-2i g \left(\cos{\theta_W} Z_\mu W^+_\nu +
\sin{\theta_W} A_\mu W^+_\nu - W^+_\mu \cos{\theta_W} Z_\nu
- W^+_\mu \sin{\theta_W} A_\nu \right) \bigg) W^-_\alpha \Bigg) u^{(i)}_L
\\ \nonumber &&
- i \frac{1}{2}
\sum_i\bar u^{(i)}_R \left(\frac{1}{2} \theta^{\mu \nu}
  \gamma^\alpha + \theta^{\nu \alpha}
  \gamma^\mu \right) \\ && \nonumber
\Bigg( \bigg(\cos{\theta_W} \partial_\mu A_\nu
-\cos{\theta_W} \partial_\nu A_\mu
-\sin{\theta_W} \partial_\mu Z_\nu
+\sin{\theta_W} \partial_\nu Z_\mu \bigg)
\\ && \nonumber
\bigg( g' Y \partial_\alpha - i Y^2 g'^2 \cos{\theta_W} A_\alpha
+ i Y^2 g'^2 \sin{\theta_W} Z_\alpha \bigg)  \Bigg) u^{(i)}_R
\\ \nonumber &&
- i \frac{1}{2}
\sum_i \bar d^{(i)}_L \left(\frac{1}{2} \theta^{\mu \nu}
  \gamma^\alpha + \theta^{\nu \alpha}
  \gamma^\mu \right) \\ && \nonumber
\Bigg( \bigg(\cos{\theta_W} \partial_\mu A_\nu
-\cos{\theta_W} \partial_\nu A_\mu
-\sin{\theta_W} \partial_\mu Z_\nu
+\sin{\theta_W} \partial_\nu Z_\mu \bigg)
\\ && \nonumber
\bigg( g' Y \partial_\alpha - i Y^2 g'^2 \cos{\theta_W} A_\alpha
+ i Y^2 g'^2 \sin{\theta_W} Z_\alpha - i \frac{1}{2}
Y g' g  \cos{\theta_W} Z_\alpha
\\ && \nonumber
- i \frac{1}{2} Y g' g  \sin{\theta_W} A_\alpha 
\bigg)
- \frac{1}{2}\bigg( \cos{\theta_W} \partial_\mu Z_\nu
 - \cos{\theta_W} \partial_\nu Z_\mu
 + \sin{\theta_W} \partial_\mu A_\nu
- \sin{\theta_W} \partial_\nu A_\mu
\\ && \nonumber
-2ig (W^+_\mu W^-_\nu - W^+_\nu W^-_\mu) \bigg) \bigg (
g \partial_\alpha -i Y g' g \cos{\theta_W}A_\alpha
\\ && \nonumber
 + i Y g' g  \cos{\theta_W} Z_\alpha
 + \frac{1}{2} i g^2 \cos{\theta_W}Z_\alpha
 + \frac{1}{2} i g^2 \sin{\theta_W}A_\alpha
\bigg)
\\ \nonumber &&
-\frac{i}{2} g^2  \bigg(\partial_\mu W^-_\nu -\partial_\nu W^-_\mu
\\ \nonumber &&
+2i g \left(\cos{\theta_W} Z_\mu W^-_\nu +
\sin{\theta_W} A_\mu W^-_\nu - W^-_\mu \cos{\theta_W} Z_\nu
- W^-_\mu \sin{\theta_W} A_\nu \right) \bigg) W^+_\alpha \Bigg) d^{(i)}_L
\\ \nonumber &&
- i \frac{1}{2}
\sum_i\bar d^{(i)}_R \left(\frac{1}{2} \theta^{\mu \nu}
  \gamma^\alpha + \theta^{\nu \alpha}
  \gamma^\mu \right) \\ && \nonumber
\Bigg( \bigg(\cos{\theta_W} \partial_\mu A_\nu
-\cos{\theta_W} \partial_\nu A_\mu
-\sin{\theta_W} \partial_\mu Z_\nu
+\sin{\theta_W} \partial_\nu Z_\mu \bigg)
\\ && \nonumber
\bigg( g' Y \partial_\alpha - i Y^2 g'^2 \cos{\theta_W} A_\alpha
+ i Y^2 g'^2 \sin{\theta_W} Z_\alpha \bigg)  \Bigg) d^{(i)}_R.
\end{eqnarray}
Note that we have not included the interactions with the gluons in
 the ``electroweak'' neutral currents.

\section*{Appendix C: Kinetic terms for the gauge bosons}

Here we will discuss the kinetic terms for the
gauge bosons in more detail and will propose an alternative
to the choice presented in the main part of the paper.

Let us reconsider the discussion of charge in non-commutative QED in
section~\ref{NCQED}. We found that without knowledge of the existence
of Seiberg-Witten
maps we would conclude that we need to  introduce a separate physical 
gauge field $\hat a^{(n)}_\mu$ 
for every charge~$q^{(n)}$ in the model. Equivalently we can also
say that the mathematical field $\widehat A_\mu$ depends nonlinearily
on the charge operator $Q$, i.e., it is enveloping algebra-valued.
Then $\widehat A_\mu  \widehat\Psi^{(n)} 
\equiv e q^{(n)} \hat a^{(n)}_\mu  \widehat\Psi^{(n)}$ with
$\hat a^{(n)}_\mu \neq \hat a^{(m)}_\mu$ for $q^{(n)} \neq q^{(m)}$.
The gauge field $\hat a^{(n)}_\mu$ appears in the covariant derivative
\eq
\widehat D_\mu \widehat \Psi^{(n)} = \pp\mu\widehat \Psi^{(n)}  
- i e q^{(n)} \hat a^{(n)}_\mu \star \widehat \Psi^{(n)}. \label{covderiv}
\en
 It is natural to provide a kinetic term for
each of these gauge fields $\hat a^{(n)}_\mu$, i.e.,
\eq
S_\mathrm{NCQED} = -\frac{1}{4N} \int d^4 x \, \sum_{n=1}^N \hat f^{(n)}_{\mu\nu} \star \hat f^{(n)}{}^{\mu\nu}  
\quad (+ \mathrm{fermions}),
\en
where the field strength $\hat f^{(n)}_{\mu\nu}$ corresponding to the gauge field $\hat a^{(n)}_\mu$ 
is determined by
\eq
\widehat F_{\mu\nu} \widehat\Psi^{(n)} 
\equiv e q^{(n)} \hat f^{(n)}_{\mu\nu}  \widehat\Psi^{(n)}, \label{onee},
\en
with $\widehat F_{\mu\nu} = \pp\mu \widehat A_\nu - \pp\nu \widehat A_\mu - i\s[\widehat A_\mu,\widehat A_\nu]$.
The factor $1/N$ in front of the action takes care of the correct commutative limit.
We can also write the action in terms of $\widehat F_{\mu\nu}$, the charge operator $Q$ and
an appropriately normalized trace $\mbox{\bf Tr}$ over the states $\widehat \Psi^{(n)}$:
\eq
S_\mathrm{NCQED} = -\frac{1}{2}\int d^4 x \, \mbox{\bf Tr} \frac{1}{(eQ)^2} \widehat F_{\mu\nu} \star \widehat F^{\mu\nu}  
\quad (+ \mathrm{fermions}).
\en
From a physical point of view there is no reason to use the same coupling constant $e$ for all 
gauge fields $\hat a_\mu^{(n)}$ in equation~(\ref{covderiv}). We could as well introduce individual coupling
constants  
and correspondingly 
rescaled fields $\hat a'^{(n)}_{\mu}$, $\hat f'^{(n)}_{\mu\nu}$.
This leads to an alternative action
\eq
S'_\mathrm{NCQED} = -\frac{1}{2}\int d^4 x \, \mbox{\bf Tr} \frac{1}{G^2} \widehat F_{\mu\nu} \star \widehat F^{\mu\nu}  
\quad (+ \mathrm{fermions}),
\en
where $G$ is an operator that is a function of the charge operator $Q$ and certain constants $g_n$, such that
\eq
G \widehat\Psi^{(n)} \propto g_n \widehat\Psi^{(n)} \qquad \mathrm{and}
\qquad
\mbox{\bf Tr} \frac{1}{G^2} \widehat F_{\mu\nu} \star \widehat F^{\mu\nu}
=
\frac{1}{N}\sum_{n=1}^N \frac{e^2}{g_n^2}(q^{(n)})^2 \hat f'^{(n)}_{\mu\nu} \star \hat f'^{(n)}{}^{\mu\nu}.
\en
The usual coupling constant $e$ can be expressed in terms of the  $g_n$ by
\eq
\mbox{\bf Tr}  \frac{1}{G^2} Q^2 = \sum_{n=1}^N \frac{1}{g_n^2}(q^{(n)})^2 = \frac{1}{2 e^2}.
\en
In the classical limit only this combination of the $g_n$ is relevant.
 
We have chosen a set-up that can be directly generalized to more general gauge theories
including the Standard Model. The action for non-Abelian noncommutative
gauge bosons is
\eq
S_\mathrm{gauge} 
= -\frac{1}{2}\int d^4 x \, \mbox{\bf Tr} \frac{1}{\mbox{\bf G}^2} \widehat F_{\mu\nu} \star \widehat F^{\mu\nu},
\en
with the non-commutative field strength $\widehat F_{\mu\nu}$, 
an appropriate trace $\mbox{\bf Tr}$ and an 
operator~$\mbox{\bf G}$. This operator must commute with all generators ($Y$, $T_L^a$, $T_S^b$)
of the gauge group so that it does not spoil the trace property of $\mbox{\bf Tr}$.  
>From what we have discussed above, it is natural to choose a trace over all the particles
(with different quantum numbers) in the model that have covariant derivatives acting on them.
In the Standard Model these are for each generation five multiplets of fermions 
and one Higgs multiplet, see Table~1. The operator $\mbox{\bf G}$ is in general a function
of $Y$ and the casimirs of $SU(2)$ and $SU(3)$. However, due to the special asignment of hypercharges
in the Standard Model it is possible to express~$\mbox{\bf G}$ just in terms
of $Y$ and six constants $g_1$, \ldots $g_6$ corresponding to the six multiplets.
In the classical limit only certain combinations of these six constants, corresponding
to the usual coupling constants $g'$, $g$, $g_S$, are relevant. The relation is
given by the following equations:
\eq
\frac{1}{g_1^2} + \frac{1}{2 g_2^2} + \frac{4}{3 g_3^2} + \frac{1}{3 g_4^2} + \frac{1}{6 g_5^2} + \frac{1}{2 g_6^2}
= \frac{1}{2g'^2},\quad
\frac{1}{g_2^2} + \frac{3}{g_5^2} + \frac{1}{g_6^2} = \frac{1}{g^2}, \quad
\frac{1}{g_3^2} + \frac{1}{g_4^2} +\frac{2}{g_5^2} = \frac{1}{g_S^2}.
\label{simplex}
\en
These three equations define for fixed $g'$, $g$, $g_S$ a three-dimensional simplex in the six-dimensional moduli space
spanned by  $1/g_1^2$, \ldots, $1/g_6^2$. The remaining three degrees of freedom become relevant at order $\theta$
in the expansion of the non-commutative action. Interesting are in particular the following traces
corresponding to tripple gauge boson vertices:
\eq
\mbox{\bf Tr} \frac{1}{\mbox{\bf G}^2} Y^3 
= -\frac{1}{g_1^2} - \frac{1}{4 g_2^2} + \frac{8}{9 g_3^2} - \frac{1}{9 g_4^2} +
\frac{1}{36 g_5^2} + \frac{1}{4 g_6^2},
\en
\eq
\mbox{\bf Tr} \frac{1}{\mbox{\bf G}^2} Y T_L^a T_L^b 
= \frac{1}{2}\delta^{ab}\left(-\frac{1}{2g_2^2} + \frac{1}{2g_5^2} +
\frac{1}{2g_6^2} \right),
\en
\eq
\mbox{\bf Tr} \frac{1}{\mbox{\bf G}^2} Y T_S^c T_S^d = \frac{1}{2}\delta^{cd}\left(
\frac{2}{3g_3^2} - \frac{1}{3g_4^2} +\frac{1}{3g_5^2} \right).
\en
We could choose, e.g., to maximize  the traces over $Y^3$
and $Y T_L^a T_L^b$. This gives 
$1/g_1^2 = 1/(2 g'^2) - 4/(3g_S^2) -1/(2 g^2)$,  $1/g_3^2 = 1/g_S^2$, $1/g_6^2 = 1/g^2$,
$1/g_2^2 = 1/g_4^2 = 1/g_5^2 = 0$
and
\[
\mbox{\bf Tr} \frac{1}{\mbox{\bf G}^2} Y^3 = -\frac{1}{2 g'^2} + \frac{3}{4 g^2} + \frac{20}{9 g_S^2},\quad
\mbox{\bf Tr} \frac{1}{\mbox{\bf G}^2} Y T_L^a T_L^b = \frac{1}{4 g^2} \delta^{ab}, \quad
\mbox{\bf Tr} \frac{1}{\mbox{\bf G}^2} Y T_S^c T_S^d = \frac{2}{6 g_S^2}\delta^{cd}.
\]
In the scheme that we have presented in the main part of this paper all three
traces are zero.
One consequence is that while non-commutativity does not \emph{require}
a triple photon vertex, such a vertex is nevertheless consistent with  non-commutativity.
It is important to note that the values of all three traces are bounded
for any choice of constants.

\end{document}